\documentstyle[tighten,preprint,prd,aps]{revtex}
\begin{document}
\title{Analysis of $K_L \to \gamma \nu \bar{\nu}$}
\author{S. Richardson and C. Picciotto}
\address{Department  of Physics and Astronomy,
         University of Victoria, Victoria,
         British Columbia, Canada V8W 3P6}
\date{\today}
\maketitle
\begin{abstract}

The decay $K_L \to \gamma \nu \bar{\nu}$ is analyzed within the
standard model. Short-distance contributions are found to dominate,
yielding a branching ratio of $\sim 0.7 \times 10^{-11}$.  We examine
the possibility that non-standard model effects might be observed at
higher rates. As an example, we calculate the branching ratio for
the process mediated by neutral horizontal gauge bosons.
Notwithstanding some model uncertainties, experimental limits for
lepton number violation constrains these contributions to $\lesssim
2\times 10^{-11}$.

\begin{center}
(Accepted for publication in Physical Review D)
\end{center}

\end{abstract}
\narrowtext
\section{Introduction}
\label{sec:intro}

Rare kaon decays continue to be an area of active research
\cite{kRevs}.  Within the standard model, rare decays are governed by
loop level processes and thus directly probe the quantum structure of
the theory.  Experimental searches serve several complementary
purposes. Improved limits help to constrain standard-model parameters,
while discovery of clear discrepancies with standard-model predictions
would signal new physics. Proposed extensions to the standard model
are similarly constrained.

Of particular interest are rare decays which are dominated by
short-distance contributions and hence are free of theoretical
uncertainties from long-distance effects. A well-known example is $K^+
\to \pi^+ \nu \bar{\nu}$ \cite{gail74,kInam81,kPlus}.  This decay is
expected to give a clean determination of the Kobayashi-Maskawa matrix
element $V_{td}$, provided the top quark mass is well-known.  Another
example is $K_L \to \pi^0 \nu \bar{\nu}$ \cite{gail74,kZero}, which is
of special interest because its dominant contribution is expected to
arise from direct $CP$ violation.

A rare kaon decay which has not been well-studied is $K_L \to \gamma
\nu \bar{\nu}$.  Although the reasons for its study are perhaps not as
compelling as those for the previous two processes, it is just as
important to test this decay for consistency with standard-model
predictions; as noted above, rare processes with very low rates are
fertile ground in which to look for non-standard effects.
Furthermore, information about this process will be useful in
background analysis of other rare-decay experiments.  An early
calculation by Ma and Okada \cite{ma78} predicted a negligible
branching ratio of $\sim 10^{-18}$. Their calculation, however,
predates the confirmation of the 3rd quark family and clearly needs to
be updated.  In Sec.\ \ref{sec:long} we examine some long-distance
processes and find contributions up to $0.7 \times 10^{-12}$.
Short-distance effects are considered in Sec.\ \ref{sec:short}, and we
find a branching ratio of $0.7 \times 10^{-11}$.  Thus within the
standard model $K_L \to \gamma \nu \bar{\nu}$ also appears to be
short-distance dominated.

One of the defining characteristics of the standard model is the
absence of tree-level flavor-changing neutral currents (FCNC).
However, many proposed extensions (technicolor theories for example)
predict such currents.  Rare neutral current kaon decays provide an
ideal window to look for such effects. In Sec.\ \ref{sec:tech} we
consider possible FCNC contributions that could arise from neutral
generation-changing gauge bosons. Our concluding remarks appear in
Sec. \ref{sec:end}.

\section{Long-Distance} \label{sec:long}

The two long-distance contributions we consider are shown in Fig.\
\ref{fig:ld}. Momentum and polarization assignments are indicated on
the graphs. Each contribution is treated separately below. Note that
within the standard model the form of the $K_L \to \gamma \nu
\bar{\nu}$ decay amplitude is tightly constrained by Lorentz and
electromagnetic gauge invariance. Assuming $CP$ conservation, the only
allowed form is
\begin{equation}
  \varepsilon^{\mu\nu\rho\sigma} ( \bar{u} \gamma_\mu P_L v)
  k_\nu \epsilon_\rho p_\sigma,
\end{equation}
where $\bar{u}$ and $v$ are neutrino spinors, and $P_L = (1 -
\gamma^5) / 2$.

\subsection{$K_L \to P \to \gamma \nu \bar{\nu}$}

We first consider the long-distance diagrams shown in Fig.\
(\ref{fig:ld})(a). They involve a weak transition of the kaon to a
virtual pseudoscalar meson $P = \pi^0, \eta, \eta^\prime$.  The
corresponding amplitude $A$ can be written in the form
\begin{equation}
   A = {\sqrt{2}
       \langle \gamma \nu \bar{\nu} |{\cal H} | \pi^0 \rangle
       \langle \pi^0 |{\cal H} | K^0 \rangle
       \over
       m_K^2 - m_{\pi^0}^2 }
       ( 1 + A_\eta + A_{\eta^{\prime}} ),
\end{equation}
assuming $\langle P|{\cal H}| K^0 \rangle = \langle P|{\cal H}|
\bar{K}^0 \rangle$.  The quantities $A_\eta$ and $A_{\eta^{\prime}}$
represent the contributions of the $\eta$ and $\eta^{\prime}$ diagrams
relative to the pion graph amplitude.  Their kaon transition elements,
for example $\langle \eta |{\cal H} | K^0 \rangle$, can be related to
the corresponding $\pi^0$ transition element using SU(3) flavor
symmetry.  To account for interference effects, we carefully treat
symmetry breaking \cite{dono86}.  We use the mixing convention
\begin{mathletters}
\label{eq:theta}
\begin{eqnarray}
  |\eta\rangle &=& \cos\theta \, |\eta_8\rangle -
        \sin\theta \, |\eta_0\rangle,\\
  |\eta^{\prime}\rangle &=& \sin\theta \, |\eta_8\rangle +
        \cos\theta \, |\eta_0\rangle,
\end{eqnarray}
where \end{mathletters} $\eta_8$ and $\eta_0$ denote the usual SU(3)
octet and singlet states. The kaon transitions then satisfy,
\begin{mathletters}
\label{eq:su3}
\begin{eqnarray}
    {\langle \eta_8 |{\cal H} | K^0 \rangle
     \over \langle \pi^0 |{\cal H} | K^0 \rangle} &=&
     {1 \over \sqrt{3}} ( 1+ \zeta ), \\
  {\langle \eta_0 |{\cal H} | K^0 \rangle
     \over \langle \pi^0 |{\cal H} | K^0 \rangle} &=&
     -  \sqrt{{8 \over 3}} \rho,
\end{eqnarray}
where \end{mathletters} $\zeta$ and $\rho$ parameterize SU(3)
breaking.  Current algebra gives $\langle \pi^0\, |{\cal H} | K^0
\rangle \simeq 1 \times 10^{-7} m_K^2$ \cite{dono86}.  Chiral
perturbation theory to one loop predicts $\zeta \simeq 0.17$
\cite{dono86}. The singlet parameter $\rho$ and mixing angle $\theta$
are fit to experiment. We will take $\rho = 0.83$ and $\theta =
-20.6^\circ$ \cite{picc92a}.

The pseudoscalar transitions $\langle \gamma \nu \bar{\nu} |{\cal H}|
P \rangle$ can be described by constituent quark triangle diagrams
\cite{kTri}, shown in Fig. \ref{fig:tri}. The calculation mimics the
well-known triangle graph approach to $\pi^0 \to \gamma \gamma$
\cite{kDewi}.  We obtain
\begin{equation}
   \label{eq:tri}
   \langle \gamma \nu \bar{\nu} \,|{\cal H}| P \rangle
   = { e G_F \over \sqrt{2} \pi^2} I_P
   \varepsilon^{\mu\nu\rho\sigma} ( \bar{u} \gamma_\mu P_L v)
   k_\nu \epsilon_\rho p_\sigma,
\end{equation}
where the pseudoscalar and quark loop dependencies have been
collected into the coefficients $I_P$. It is convenient to express
$I_P$ in terms of the mass ratios $\xi_q \equiv p^2/ m_q^2$ for
constituent quark flavor $q$. Then
\begin{equation}
  \label{eq:ip}
  I_P = 2 N_c \sum_q Q_q
       { g_{Pq\bar{q}} \over m_q } { c_{Vq} \over \xi_q}
       \left( \arcsin \sqrt{\xi_q / 4} \right)^2,
\end{equation}
provided $0 \leq \xi_q \leq 4$. Here $N_c = 3$ is the number of
colors, $Q_q$ are the quark charges in units of $e$, and $c_{Vq}$ are
the standard-model weak vector coupling coefficients.  The effective
meson-quark couplings $g_{Pq\bar{q}}$ are shown in Table
\ref{tab:su3}.  The parameters $f_8 = 1.25 f_\pi$ and $f_0= 1.04
f_\pi$, where $f_\pi = 132$ MeV, account for SU(3) breaking in the
$\eta_8$ and $\eta_0$ amplitudes \cite{dono85}.  For the constituent
quark masses, we choose the representative values $m_u = m_d = 0.31$
GeV and $m_s = 0.46$ GeV.  The resulting coefficient values are
$I_{\pi^0} = 0.25$, $I_{\eta_8} = 0.51$, and $I_{\eta_0} = 5.7$
$\text{GeV}^{-1}$.

In terms of Eqs. (\ref{eq:theta}), (\ref{eq:su3}) and (\ref{eq:ip}),
the quantities $A_\eta$ and $A_{\eta^{\prime}}$ are
\begin{mathletters}
\label{eq:aa}
\begin{eqnarray}
   A_\eta &=&
      {m_K^2 - m_{\pi^0}^2 \over m_K^2 - m_\eta^2}
      \biggl(
         { I_{\eta_8} \over I_{\pi^0} } \cos\theta
         - { I_{\eta_0} \over I_{\pi^0} } \sin\theta
      \biggr) \nonumber \\
      && \times \biggl(
         { 1+\zeta \over \sqrt{3} } \cos\theta
         + \sqrt{{ 8 \over 3}}\rho \, \sin\theta
      \biggr), \\
   A_{\eta^\prime} &=&
      {m_K^2 - m_{\pi^0}^2 \over m_K^2 - m_{\eta^\prime}^2}
      \biggl(
         { I_{\eta_8} \over I_{\pi^0} } \sin\theta
         +  { I_{\eta_0} \over I_{\pi^0} } \cos\theta
      \biggr) \nonumber\\
      && \times \biggl(
         { 1+\zeta \over \sqrt{3} }  \sin\theta
         - \sqrt{{ 8 \over 3}}\rho \, \sin\theta
      \biggr).
\end{eqnarray}
Inserting \end{mathletters} the numerical values for $I_P$ gives
$A_\eta = -6.9$ and $A_{\eta^\prime} = 10.6$.  Integrating the
amplitude $A$ over phase space gives the corresponding partial rate
$\Gamma$ for a single neutrino family
\begin{equation}
   \Gamma =
   { \alpha G^2_F I_{\pi^0}^2\, m_{\pi^0}^7 \over 60 (2 \pi)^6}
   { |\langle \pi^0 |{\cal H} | K^0 \rangle|^2
        \over (m_K^2 - m_{\pi^0}^2)^2 }
   (1 + A_\eta + A_{\eta^{\prime}})^2,
   \label{eq:lda}
\end{equation}
where $\alpha = e^2 / 4 \pi$ is the fine structure constant.  For
three neutrino families, we find using Eq.\ (\ref{eq:lda}) a branching
ratio of $0.8 \times 10^{-17}$.

\subsection{$K_L \to K^*(\gamma) \to \nu \bar{\nu}$}

The second long-distance contribution we consider is shown in Fig.\
(\ref{fig:ld})(b). The diagram involves a radiative transition of the
kaon to an intermediate $K^*$. The corresponding amplitude $A$ takes
the form
\begin{equation}
   A = { (\langle\nu \bar{\nu} |{\cal H}| K^{*0} \rangle
         + \langle\nu \bar{\nu} |{\cal H}|\bar{K}^{*0} \rangle)
       \cdot \langle K^{*0} \gamma |{\cal H}| K^0 \rangle
       \over \sqrt{2} (q^2 - m_{K^*}^2) },
   \label{eq:form}
\end{equation}
assuming $\langle \bar{K}^{*0} \gamma |{\cal H}| \bar{K}^0\rangle =
\langle K^{*0} \gamma |{\cal H}| K^0\rangle$. To model the photon
vertex, we normalize the amplitude predicted by the vector meson
dominance hypothesis to the experimental rate for $K^* \to K^0
\gamma$.  The amplitude is \cite{kDurso}
\begin{equation}
  A_{K^* \to K^0 \gamma} = e g
      \varepsilon_{\mu\nu\rho\sigma}
      q^\mu E^\nu k^\rho \epsilon^\sigma,
   \label{eq:one}
\end{equation}
where $g$ is the effective vertex coupling.  The corresponding decay
rate is
\begin{equation}
   \Gamma_{K^* \to K^0 \gamma} =
   {\alpha |g|^2 m_{K^*}^3 \over 24}
   \biggl( 1- { m_{K^0}^2 \over m_{K^*}^2 } \biggr)^3.
\end{equation}
The measured branching ratio for $K^* \to K^0 \gamma$ is $2.3 \times
10^{-3}$ \cite{pdg94}. This implies $|g| = 1.3$ $\text{GeV}^{-1}$.

The $K^* \to \nu \bar{\nu}$ transition is modelled as a short-distance
process.  In the standard model, the quark-level interaction $d\bar{s}
\to \nu \bar{\nu}$ proceeds at the one-loop level. As indicated in
Fig.  \ref{fig:loop}, both box and effective $dsZ$ vertex diagrams
contribute. The corresponding effective four-fermion interaction is
\cite{kInam81}
\begin{equation}
  \label{eq:Li}
  {\cal H}_i = 2 \sqrt{2} G_{\text{F}} \chi\tilde{D}_i
  \, \bar{s} \gamma_\mu P_L d
  \,\, \bar{\nu}_{i} \gamma^\mu P_L \nu_{i}
  + \mbox{h.c.},
\end{equation}
where $\chi = \alpha / 4 \pi \sin^2 \theta_W$ and $i = e,\mu,\tau$
denotes the neutrino family.  The quantity $\tilde{D}_i$ is a function
of the appropriate lepton mass, the up-type quark masses, and several
elements of the Kobayashi-Maskawa matrix. Our determinations for
$\tilde{D}_i$ are described in the Appendix.  The $K^*$ transitions
determined by ${\cal H}_i$ contain the hadronic matrix element
$\langle0|\bar{s} \gamma_\mu P_L d| K^*\rangle \equiv f_{K^*}E_\mu$.
We will take $f_{K^*} = f_\rho = (0.2) m_\rho^2$ \cite{kDGH}.

Eqs. (\ref{eq:one}) and (\ref{eq:Li}) determine all the
transition elements appearing in Eq.\ (\ref{eq:form}).  The resulting
amplitudes  $A_i$ for a single neutrino family are
\begin{equation}
   A_i = {4 e g G_F \chi f_{K^*} \text{Re} \tilde{D}_i
          \over q^2 - m_{K^*}^2 }
  \varepsilon^{\mu\nu\rho\sigma} ( \bar{u}_i \gamma_\mu P_L v_i)
  k_\nu \epsilon_\rho p_\sigma.
\end{equation}
The corresponding partial decay rates are
\begin{equation}
   \Gamma_i = {\alpha |g|^2
               (G_F \chi f_{K^*} \text{Re} \tilde{D}_i)^2
               m_{K}^3 \over 12 \pi^2} F(x),
\end{equation}
where $x = m_{K^*}^2 / m_{K}^2$ and $F(x)$ is a kinematic
function. Specifically,
\[
   F(x) = -\frac{17}{6} + 7x - 4x^2 - (x-1)^2(4x-1)\ln {x-1 \over x}.
\]
Summing over the three neutrino families gives $\Gamma= 8.5 \times
10^{-30}$ GeV, or a branching ratio of $0.7 \times 10^{-12}$.

\section{Short-distance}
\label{sec:short}

For $K_L \to \gamma \nu \bar{\nu}$, the relevant short-distance
interaction is $d\bar{s} \to \gamma \nu \bar{\nu}$.  To classify the
lowest order diagrams, it is convenient to start with the graphs for
$d\bar{s} \to \nu \bar{\nu}$. These were shown earlier in Fig.\
\ref{fig:loop}. Adding a photon to any charged line creates a graph
for $d\bar{s} \to \gamma \nu \bar{\nu}$. Provided the standard-model
$WWZ\gamma$ vertex is included, this method generates the full set of
lowest order diagrams. The short-distance contributions thus naturally
split into box (Fig.\ \ref{fig:box}) and effective $dsZ$ type (Fig.\
\ref{fig:dsZ}) diagrams. Each class further divides into one-particle
irreducible and one-quark reducible processes.

Ma and Okada \cite{ma78} only considered the leading contribution to
the effective $dsZ$ type diagrams: the one-particle irreducible graphs
containing a single $W$ boson \cite{gail74}.  Assuming just two quark
families, they found the following short-distance amplitude for $K_L
\to \gamma \nu \bar{\nu}$:
\begin{equation}
   {\sqrt{2} e G_F^2 \sin \theta_c \cos \theta_c c_{Vu} f_K
    \over 3 \pi^2 }
  \varepsilon^{\mu\nu\rho\sigma} ( \bar{u} \gamma_\mu P_L v)
  k_\nu \epsilon_\rho p_\sigma,
\end{equation}
where $\theta_c$ is the Cabbibo angle. Their result was dominated by
the graphs containing an intermediate $u$ quark. The heavier charm
quark contributions were shown to be small. The analysis, however, did
not take into account the possibility of an intermediate quark mass
larger than the $W$ mass.

We have extended the Ma and Okada calculation to the case of three
fermion families, and a large top quark mass. In the large mass
approximation, the above amplitude would be modified by the factor
\[
  {p\cdot k \over 6 m_W^2}
  {V_{td} V_{ts}^* \over \sin \theta_c \cos \theta_c}
  \biggl[ {2\over (z-1)^2} - {2 \ln z \over (z-1)^3} - {1\over z(z-1)}
  \biggr],
\]
where $z=m_t^2 / m_W^2$.  For $m_t \approx 174$ GeV \cite{cdf94}, this
factor actually reduces the corresponding rate by several orders of
magnitude. Thus Ma and Okada's conclusion that the effective $dsZ$
diagrams are dominated by the lightest quark is still valid.

Ma and Okada, however, did not consider the box diagram contributions.
Our expectation is that the major short-distance contributions in fact
come from the one-quark reducible box diagrams in Fig.\
\ref{fig:box}(b).

In order to later treat non-standard model effects, we approach the
calculation in a relatively general manner. As indicated in Fig.\
\ref{fig:sd}, we consider all one-quark reducible processes that can
be viewed as a photon attached to some effective $d\bar{s}\to
\nu_j\bar{\nu}_k$ interaction. The form of the four-fermion
Hamiltonian is taken to be
\begin{equation}
   \label{eq:heff}
   {\cal H} = \beta_{jk} \,
              \bar{s} D^\mu d \, \bar{\nu}_j N_\mu \nu_k
              + \text{h.c.},
\end{equation}
where $D_\mu \equiv \gamma_\mu ({\sf v}_D - {\sf a}_D \gamma_5)$ and
$N_\mu \equiv \gamma_\mu ( {\sf v}_N - {\sf a}_N \gamma_5)$ are
arbitrary mixtures of vector and axial-vector components. The
standard-model effective four-fermion Hamiltonian was discussed
previously in Eq.\ (\ref{eq:Li}). In this case, the fermion currents
are both left-handed, lepton number conservation implies $\beta_{j\neq
  k} = 0$, and $\beta_{jj}= 2\sqrt{2} G_F \chi \tilde{D}_j$. Note that
by using these coefficients, several effective $dsZ$ type graphs will
be included along with the one-quark reducible box diagrams.

Using the interaction ${\cal H}$, the quark-level amplitude generated
by the processes in Fig.\ \ref{fig:sd} is
\begin{eqnarray}
   A = {e \beta_{jk} \over 6} \bar{v}_s & & \biggl[
   { D^\mu (\not\!p_d - \not\!k + m_d) \not\!\epsilon
       \over p_d \cdot k}
   \nonumber \\ & & \;\;\;
   + { \not\!\epsilon (\not\!p_s - \not\!k + m_s) D^\mu
       \over p_s \cdot k}
   \biggr] u_d \,
   \bar{u}_j N_\mu v_k,
\end{eqnarray}
where $m_d$ and $m_s$ are quark current masses.  It remains to extract
the corresponding kaon decay amplitude.  For a reasonable estimate we
project out the ``pseudoscalar'' content of $A$ with the operation
\cite{gold77}
\begin{equation}
   \label{eq:pro}
   M = {\cal C}  \sum \bar{u}_s \gamma^5 v_d \, A,
\end{equation}
where the sum is over the quark spins and ${\cal C}$ is a
normalization constant. Only terms that conserve $CP$ are retained.
In addition a simple static constituent quark model is used to
evaluate the quark momenta. The resulting projected amplitude for $K^0
\to \gamma \nu_j \bar{\nu}_k$ is
\begin{equation}
   M =  { 4 {\cal C} e \beta_{jk} \over 3 p \cdot k}
   (m_s - m_d) {\sf v}_D \varepsilon^{\mu\rho\nu\sigma}
   (\bar{\nu}_j N_\mu \nu_k)  k_\rho \epsilon_\nu p_\sigma
\end{equation}
where $m_d$ and $m_s$ are now constituent quark masses. The
$\bar{K^0}$ amplitude is identical except that $\beta_{jk}^*$ appears.
Notice that only the vector part of the quark current contributes.

The normalization constant ${\cal C}$ is determined by applying the
projection to the decay $K^+ \to l^+ \nu_l$. Comparison with the
definition of the kaon decay constant gives ${\cal C} = f_K m_K / 8
m_s m_d V_{us}^*$.

After combining the $K^0$ and $\bar{K}^0$ amplitudes, and integrating
over phase space, we obtain the corresponding $K_L \to \gamma \nu
\bar{\nu}$ decay rate
\begin{equation}
  \label{eq:sdrate}
  \Gamma = {\alpha \over 2 (72 \pi)^2}
           ({\sf v}_N^2 + {\sf a}_N^2) {\sf v}_D^2
           \text{Tr}[(\beta + \beta^\dagger)^2]
           { f_K^2 \rho^2 m_K^3 \over |V_{us}|^2 },
\end{equation}
where $\rho \equiv m_s /m_d - m_d / m_s$.  Inserting the
standard-model couplings and coefficient matrix $\beta$ gives
\begin{equation}
   \Gamma = {2 \alpha G_F^2 \chi^2 \over (72 \pi)^2 }
   \sum_i (\text{Re} \tilde{D}_i)^2
   { f_K^2 \rho^2 m_K^3 \over |V_{us}|^2 }.
\end{equation}
Our numerical evaluation of the Inami and Lim function $\tilde{D}_i$
is discussed in the Appendix.  Estimates for the constituent quark
masses vary substantially.  We will take $\rho = 0.8 \pm 0.2$ to
account for some of this uncertainty.  The remaining parameters are
all well-known \cite{pdg94}. The resulting branching ratio is
$0.7^{+0.6}_{-0.4} \times 10^{-11}$.

\section{Horizontal Gauge Bosons}
\label{sec:tech}

Since $K_L \to \gamma \nu \bar{\nu}$ is highly suppressed in the
standard model, observation of this decay at a higher rate would
signal new physics. Therefore, it is of interest to look at the
predictions of extended models. One possibility is the contribution of
flavor-changing neutral currents (FCNC).  Here we consider the effects
of neutral generation-changing gauge bosons and, in particular, we
follow the general formalism adopted by Cahn and Harari \cite{cahn80}.
We have extended this formalism to three generations, as follows.

The general group structure is taken to be $G \times H$, where $G$
acts within a generation and $H$ is a ``horizontal'' gauge group
containing neutral generation-changing bosons. The group $G$ contains
the standard electroweak algebra $\text{SU(2)}_W \times \text{U(1)}$.
The group $H$ is taken to be $\text{SU(2)}_H$. We will further assume
that the three horizontal gauge bosons are degenerate in mass.

The basis of particle states is constructed as follows. Let $L^0_1$,
$L^0_2$ and $L^0_3$ denote the ``primitive'' electron, muon, and tau.
For the present, it is convenient to ignore the distinction between
left and right components. These primitive states are eigenstates of
both $\text{SU(2)}_W$ and $\text{SU(2)}_H$ with eigenvalues $T^W_3 =
-1/2$ and $T^H_3 = -1,0,1$ respectively. Similarly, let the triplets
$N^0_i$, $U^0_i$ and $D^0_i$ denote the primitive states of the
neutrinos, up-type, and down-type quarks.

The set of mass eigenstates is assumed to be related to the primitive
states by a unitary transformation. Specifically,
\[
   F \equiv
     \left[ \begin{array}{c} N \\ L \\ U \\ D \end{array} \right]
   = \left[ \begin{array}{cccc}
             {\cal U}^N &&& \\
            & {\cal U}^L && \\
            && {\cal U}^U & \\
            &&& {\cal U}^D
            \end{array} \right]
     \left[ \begin{array}{c}
            N^0 \\ L^0 \\ U^0 \\ D^0 \end{array} \right]
   \equiv {\cal U} F^0.
\]
In the primitive basis, the horizontal currents have the simple
representation $\bar{F}^0 \bbox{T}^H F^0$ where $\bbox{T}^H =
\openone_4 \otimes \bbox{\tau}$ and the $\tau_i$ are the generators
for the 3-dimensional representation of SU(2).  For convenience the
Lorentz structure of the currents has been temporarily suppressed.
The horizontal gauge bosons thus presumably give rise to the following
low-energy effective interaction
\begin{equation}
   \label{eq:lAll}
   {\cal H} = {g_H^2 \over 2 m_H^2}
       (\bar{F} {\cal U} \bbox{\tau} {\cal U}^\dagger F)^2,
\end{equation}
where $g_H$ is the gauge coupling constant and $m_H$ is the
mass of the gauge bosons.

For the decay $K_L \to \gamma \nu \bar{\nu}$, the relevant
interactions in ${\cal H}$ are those that connect the $D$ and
$N$ triplets. These can be written as
\begin{equation}
    {\cal H}_{DN} = {g_H^2 \over m_H^2}
       \bar{D} \bbox{\tau} D \cdot
       \bar{N} {\cal U}^N {\cal U}^{D\dagger} \bbox{\tau}
               {\cal U}^{D} {\cal U}^{N\dagger} N,
\end{equation}
by making the replacement $\bbox{\tau} \to {\cal U}^{D\dagger}
\bbox{\tau} {\cal U}^{D}$. Furthermore, keeping just the terms
involving the $d$ to $s$ quark transitions gives
\begin{equation}
   {\cal H}_{ds} = {g_H^2 \over \sqrt{2} m_H^2}
       ({\cal U}^N {\cal U}^{D\dagger} \tau_-
        {\cal U}^D {\cal U}^{N\dagger})_{jk} \,
        \bar{s}d \, \bar{\nu}_j \nu_k  + \text{h.c.}.
\end{equation}
Comparison of this expression with the general Hamiltonian given in
Eq.\ (\ref{eq:heff}) determines the coefficients $\beta_{jk}$ that
appeared in our short-distance analysis.  Thus the decay rate given in
Eq.\ (\ref{eq:sdrate}) can be used directly.  Note that the unitary
mixing matrices do not contribute to the trace.  The resulting $K_L
\to \gamma \nu \bar{\nu}$ rate is
\begin{equation}
   \label{eq:rateH}
     \Gamma = {2 \alpha \over (72 \pi)^2}
           ({\sf v}_N^2 + {\sf a}_N^2) {\sf v}_D^2
           {g_H^4 \over m_H^4}
           { f_K^2 \rho^2 m_K^3 \over |V_{us}|^2 }.
\end{equation}
The neutral generation-changing bosons that we are considering also
mediate the lepton-number violating processes $K_L \to e \mu$ and $K^+
\to \pi^+ e \mu$. The stringent experimental limits on these processes
help to constrain the horizontal gauge boson couplings and mass.  The
process $K_L \to e \mu$, however, is only sensitive to the
axial-vector component of the quark currents.  Thus the semi-leptonic
decay $K^+ \to \pi^+ e \mu$, which is sensitive to the vector
component, is more convenient as a comparison process.

The relevant quark-level process for $K^+ \to \pi^+ e^- \mu^+$ is
$d\bar{s} \to e^- \mu^+$.  Repeating the steps for the $DL$
interactions that were used for the $DN$ interactions, one finds that
the appropriate term in the Hamiltonian is
\begin{equation}
   {\cal H}_{\bar{s}d\bar{e}\mu} =  {g_H^2 \over m_H^2}
       ({\cal V}_{e\mu} {\cal V}_{\mu\tau}^*
        + {\cal V}_{ee} {\cal V}_{\mu\mu}^*)
        \bar{s}d \, \bar{e} \mu,
\end{equation}
where ${\cal V} \equiv {\cal U}^L {\cal U}^{D\dagger}$ governs the
mixing between lepton and down-type quark states.  Normalizing the
corresponding decay rate to the standard-model decay $K^+ \to \pi^0
\nu_\mu \mu^+$ we find the ratio
\[
   R \equiv { \Gamma_{\pi^+ e^- \mu^+} \over
     \Gamma_{\pi^0 \nu_\mu \mu^+} }
   = ({\sf v}_L^2 + {\sf a}_L^2) {\sf v}_D^2
       { g_H^4\over m_H^4}
       { |{\cal V}_{e\mu} {\cal V}_{\mu\tau}^*
         + {\cal V}_{ee} {\cal V}_{\mu\mu}^*|^2 \over
         G_F^2|V_{us}|^2 }.
\]
The present experimental limit is $R < 6.6 \times 10^{-9}$
\cite{pdg94}.

The $K_L \to \gamma \nu \bar{\nu}$ rate appearing in Eq.\
(\ref{eq:rateH}) can be written in terms of $R$ as
\begin{equation}
   \Gamma = {\alpha G_F^2 \over (72 \pi)^2}
            { {\sf v}_N^2 + {\sf a}_N^2 \over
              {\sf v}_L^2 + {\sf a}_L^2 }
            { f_K^2 \rho^2 m_K^3 \over
            |{\cal V}_{e\mu} {\cal V}_{\mu\tau}^*
              + {\cal V}_{ee} {\cal V}_{\mu\mu}^*|^2 }
            R.
\end{equation}
In order to obtain a reasonable estimate, we ignore the remaining
dependence on horizontal coupling coefficients and mixing angles.  We
then obtain for $K_L \to \gamma \nu \bar{\nu}$ a branching ratio limit
of $\lesssim 2 \times 10^{-11}$.

\section{Concluding Remarks}
\label{sec:end}

Within the standard model, the decay $K_L \to \gamma \nu \bar{\nu}$ is
indeed very rare, but not nearly to the extent predicted by earlier
calculations \cite{ma78}.  We have found long-distance contributions
with branching ratios up to $0.7 \times 10^{-12}$ and short-distance
contributions of $0.7^{+0.6}_{-0.4} \times 10^{-11}$. We also
considered the possible effects of neutral generation-changing gauge
bosons. Present experimental limits for lepton violation suggest that
these contributions are limited to $\lesssim 2 \times 10^{-11}$.

\section{Acknowledgment}
\label{sec:ack}

This work was supported in part by the Natural Science and
Engineering Council of Canada.

\appendix
\section*{}

This appendix discusses the numerical evaluation of the Inami-Lim
function $\tilde{D}_i$ \cite{kInam81}.  Values for quantities not
specified in the text are taken from the 1994 Review of Particle
Properties \cite{pdg94}.

\subsection*{Constraints on the KM Matrix}

Since $\tilde{D}_i$ depends heavily upon the elements of the
Kobayashi-Maskawa (KM) matrix, we first update their values
\cite{kApp}.  In the Wolfenstein convention \cite{wolf83}, to order
$\lambda^3$, the KM matrix $V$ takes the form
\begin{equation}
   \label{eq:wolf}
   V = \left (
      \begin{array}{ccc}
         1 - \lambda^2 / 2 & \lambda & A \lambda^3 (\rho - i \eta) \\
         - \lambda & 1 - \lambda^2 / 2 & A \lambda^2 \\
         A \lambda^3 (1-\rho - i \eta) &- A \lambda^2 & 1
      \end{array}
   \right ).
\end{equation}
The measured element magnitudes give $\lambda = 0.221 \pm 0.002$ and
$A = 0.82 \pm 0.10$. Furthermore, the ratio $|V_{ub}/V_{cb}|$
describes a circle in the $(\rho,\eta)$ plane,
\begin{equation}
   \label{eq:con1}
   \rho^2 + \eta^2 = (0.36 \pm 0.09)^2.
\end{equation}
Two additional constraints on $\rho$ and $\eta$, as reviewed below,
are obtained from measurements of $K^0$--$\bar{K}^0$ and
$B^0_d$--$\bar{B}^0_d$ mixing.

In the neutral $B_d$ meson system, mixing is characterized by the
parameter ${\sf x}_d$.  The dominant processes are box diagrams
containing top quarks. One obtains \cite{alta88}
\begin{equation}
   \label{eq:xd}
   {\sf x}_d \simeq { G_F^2 m_W^2 \over 6 \pi^2 }
   f_B^2 B_B m_B
   \tau_B |\bar{E}(x_t)| \eta_B
   |V_{tb} V_{td}|^2,
\end{equation}
where the function $\bar{E}(x)$ denotes $\bar{E}(x,x)$, and
\cite{kInam81}
\begin{eqnarray}
   \bar{E}(x,y) = {x y \over 4} & &
   \left[
      { y^2 - 8 y + 4 \over (y - 1)^2 (x-y) } \ln y
   \right.
   \nonumber \\ & & \;\;\;
   \left.
     + { 3 \over 2(x-1)(y-1) } + (x \leftrightarrow y)
   \right].
\end{eqnarray}
The arguments $x_q = m_q^2 /m_W^2$ where $m_q$ are the quark masses.
In the Wolfenstein convention, Eq.\ (\ref{eq:xd}) describes a second
circle in the $(\rho,\eta)$ plane,
\begin{equation}
   \label{eq:con2}
   (1-\rho)^2 + \eta^2 = (1.19 \pm 0.29)^2,
\end{equation}
where we used $f_B=0.140 \pm 0.025$ GeV and $B_B=0.85 \pm 0.10$
\cite{kim90} for the hadronic parameters, $m_t= 174 \pm 16$ GeV
\cite{cdf94} for the top mass, and $\eta_B=0.85 \pm 0.05$ \cite{kim90}
for the QCD correction.

In the neutral kaon system, mixing is characterized by the $CP$
violation parameter $\epsilon$. Both charm and top quark box diagrams
contribute. One finds \cite{pasc89}
\[
   |\epsilon| \simeq
   { G_F^2 m_W^2 \over 12 \pi^2 }
   { f_K^2 B_K m_K \over \sqrt{2} \, \Delta m_K}
    | \sum_{\stackrel{\scriptstyle i=c,t}{j= c,t}}
      \bar{E}(x_i, x_j) \eta_{ij}
      \text{Im} \, \Lambda_i \Lambda_j|,
\]
where $\Lambda_q$ denotes the product $V_{qd} V_{qs}^*$. Note that
when evaluating the $\Lambda_q$, the KM matrix elements must be
expanded to order $\lambda^5$.  Using $B_K=0.7 \pm 0.2$ and $f_K =
0.160$ GeV \cite{bela91} for the hadronic factors gives the hyperbola
\begin{equation}
   \label{eq:con3}
   \eta = {
      (7.3 \pm 2.1) \times 10^{-4}
      \over
      A^2 | \eta_{cc} \bar{E}(x_c) - \eta_{ct} \bar{E}(x_c,x_t)
            - A^2 \lambda^4 (1-\rho) \eta_{tt} \bar{E}(x_t) |
   }.
\end{equation}
For calculations, the charm mass was taken to be $m_c= 1.5 \pm 0.1$
GeV, and we used $\eta_{cc}=0.76$, $\eta_{ct}=0.36$, and
$\eta_{tt}=0.61$ \cite{bela91} for the QCD corrections.

The one-sigma contours for the constraints (\ref{eq:con1}),
(\ref{eq:con2}), and (\ref{eq:con3}) are plotted in Fig.
\ref{fig:con}.  The product of these distributions determines our
expected region for $\rho$ and $\eta$. Contours for this region are
shown in Fig. \ref{fig:reg}.

\subsection*{Evaluation of $\tilde{D}_i$}

In the Wolfenstein convention for the KM matrix, the function
$\tilde{D}_i$, $i \in \{ e, \mu,\tau \}$, becomes
\[
   \tilde{D}_i
   \simeq -\lambda \eta_c \bar{D}(x_c,y_i) - \lambda^5 \eta_t A^2
                (1-\rho-i\eta) \bar{D}(x_t,y_i),
\]
where $\eta_c$ and $\eta_t$ are QCD corrections and
\begin{eqnarray}
  \bar{D}(x,y) &=& { x \over 8} \biggl\{
  \biggl[
      {y \over x-y} \biggl( {y-4 \over y-1} \biggr)^2 \ln y
    + (x \leftrightarrow y)
  \biggr] \\
  & & + 2 - 3 {y+2 \over (y-1)(x-1)}
    + {x^2 -2 x + 4 \over (x-1)^2} \ln x
  \biggr\} \nonumber
\end{eqnarray}
For the standard-model masses $y \ll 1$, and $y \ll x$ except when $y
= y_{\tau}$ and $x = x_c$. It follows that $\tilde{D}_i$ is relatively
insensitive to $i$. We again take $m_c= 1.5 \pm 0.1$ GeV and $m_t= 174
\pm 16$ GeV, and set $\eta_c = 0.7$ and $\eta_t = 1$ \cite{buch94}.
Then using the Wolfenstein parameter values determined above, we find
\begin{eqnarray}
   \label{eq:dti}
   \text{Re} \tilde{D}_{e,\mu} &=& -1.8^{+0.5}_{-0.3} \times 10^{-3},
   \nonumber \\
   \text{Re} \tilde{D}_\tau &=& -1.6^{+0.6}_{-0.3} \times 10^{-3}, \\
   \text{Im} \tilde{D}_i &=& 3.0^{+2.0}_{-1.3} \times 10^{-4}
   \nonumber.
\end{eqnarray}
The listed values are the modes and the uncertainties
enclose 68.3\% of the distributions.

\begin{figure}
   \caption{Long-distance contributions.}
   \label{fig:ld}
\end{figure}

\begin{figure}[h]
   \caption{
      Triangle diagram for the pseudoscalar decays
      $P \to \gamma \nu \bar{\nu}$. The effect of the diagram
      with the photon and $Z$ lines interchanged is to double the
      weak vector couplings $c_{Vq}$ and cancel the
      weak axial-vector couplings $c_{Aq}$.}
   \label{fig:tri}
\end{figure}

\begin{figure}
   \caption{Diagrams for $d\bar{s}\to \nu \bar{\nu}$. The four
   effective $dsZ$ vertex diagrams can be classified as either (a)
   one-particle irreducible or (b) one-quark reducible. Only one of
   each pair is drawn. (c) The box diagram.}
   \label{fig:loop}
\end{figure}

\begin{figure}
   \caption{The box diagrams for $d\bar{s}\to \gamma \nu \bar{\nu}$
   can be generated from the box diagram for $d\bar{s}\to \nu
   \bar{\nu}$ by photon emission from all charged lines.
   Shown here are examples of (a) one-particle irreducible and (b)
   one-quark reducible graphs.}
   \label{fig:box}
\end{figure}

\begin{figure}
   \caption{Effective $dsZ$ diagrams for $d\bar{s}\to \gamma \nu
   \bar{\nu}$ can be generated from the effective $dsZ$ diagrams
   for $d\bar{s}\to \nu \bar{\nu}$ by photon emission from all
   charged lines. Shown here are examples of (a) one-particle
   irreducible and (b) one-quark reducible graphs.}
   \label{fig:dsZ}
\end{figure}

\begin{figure}
   \caption{Effective short-distance processes.}
   \label{fig:sd}
\end{figure}

\begin{figure}
   \caption{One-sigma constraint curves for $\rho$ and $\eta$.
            The dashed and dotted lines respectively are the
            $V_{ub}/V_{cb}$ and $B$ mixing circles. The solid lines
            are generated from the $K$ mixing hyperbola by a Monte
            Carlo treatment of the uncertainties.}
   \label{fig:con}
\end{figure}

\begin{figure}
   \caption{Expected $(\rho,\eta)$ region. Dashed contour encloses
            68.3\% of the distribution; solid contour encloses 90\%.}
   \label{fig:reg}
\end{figure}

\begin{table}
   \caption{Effective meson-quark couplings.}
   \label{tab:su3}
   \begin{tabular}{cccc}
   $P$ & $g_{Pu\bar{u}} / m_u$
      & $g_{Pd\bar{d}} / m_d$
      & $g_{Ps\bar{s}} / m_s$ \\
   \hline
   $\pi^0$  & $(2)^{1/2} f^{-1}_\pi$ & $ - (2)^{1/2} f^{-1}_\pi$ &0\\
   $\eta_8$ & $(2/3)^{1/2} f^{-1}_8 $
      & $(2/3)^{1/2} f^{-1}_8$
      & ${ -(8/3)^{1/2} f^{-1}_8 }$ \\
   $\eta_0$ & $2 (3)^{-1/2} f^{-1}_0$
      & $2 (3)^{-1/2} f^{-1}_0$ & $2 (3)^{-1/2} f^{-1}_0$
   \end{tabular}
\end{table}


\begin{references}

\bibitem{kRevs} For reviews, see J.~L. Ritchie and S.~G. Wojcicki,
  Rev. Mod. Phys. {\bf 65}, 1149 (1993); L. Littenberg and G.
  Valencia, Annu. Rev. Nucl. Part. Sci. {\bf 43}, 729 (1993); R.
  Battiston, D. Cocolicchio, G.~L. Fogli, and N. Paver, Phys. Rep.
  {\bf 214}, 293 (1992); J.~S. Hagelin and L.~S. Littenberg, Prog.
  Nucl. Part. Phys. {\bf 23}, 1 (1989).

\bibitem{gail74} M.~K. Gaillard and B.~W. Lee, Phys. Rev. D {\bf 10},
  897 (1974).

\bibitem{kInam81} T. Inami and C.~S. Lim, Prog. Theor. Phys. {\bf 65},
  297 (1981); {\bf 65} 1772(E) (1982).

\bibitem{kPlus} M.~S. Atiya {\em et. al.}, Phys. Rev. D {\bf 48}, 1
  (1993); D. Rein and L.~M.  Sehgal, Phys. Rev. D {\bf 39}, 3325
  (1989); J. Ellis, J.~S. Hagelin, and S.  Rudaz, Phys. Lett. B {\bf
    192}, 201 (1987).

\bibitem{kZero} L.~S. Littenberg, Phys. Rev. D {\bf 39}, 3322 (1989);
  J. Ellis, M.~K. Gaillard, and D.~V. Nanopoulos, Nucl. Phys. {\bf
    B109}, 213 (1976).

\bibitem{ma78} E. Ma and J. Okada, Phys. Rev. D {\bf 18}, 4219 (1978).

\bibitem{dono86} J.~F. Donoghue, B.~R. Holstein, and Y.-C.~R. Lin,
  Nucl. Phys. {\bf B277}, 651 (1986).

\bibitem{picc92a} C. Picciotto, Phys. Rev. D {\bf 45}, 1569 (1992).

\bibitem{kTri} D. Grasso and M. Lusignoli, Phys. Lett. B {\bf 279},
  161 (1992); L. Arnellos, W.~J. Marciano, and Z. Parsa, Nucl. Phys.
  {\bf B196}, 365 (1982).

\bibitem{kDewi} For example, see B.~de. Wit and J. Smith, {\em Field
    Theory in Particle Physics} (North-Holland Physics Publishing, New
  York, 1986).

\bibitem{dono85} J.~F. Donoghue, B.~R. Holstein, and Y.-C.~R. Lin,
  Phys. Rev. Lett. {\bf 55}, 2766 (1985).

\bibitem{kDurso} For example, see J.~W. Durso, Phys. Lett. B {\bf
    184}, 348 (1987).

\bibitem{pdg94} {Particle Data Group}, Phys. Rev. D {\bf 50}, 1173
  (1994).

\bibitem{kDGH} For example, see J.~F. Donoghue, E. Golowich, and B.~R.
  Holstein, {\em Dynamics of the Standard Model} (Cambridge University
  Press, Cambridge, 1992).

\bibitem{cdf94} {CDF Collaboration}, Phys. Rev. Lett. {\bf 73}, 225
  (1994).

\bibitem{gold77} T. Goldman and W.~J. Wilson, Phys. Rev. D {\bf 15},
  709 (1977).

\bibitem{cahn80}
R.~N. Cahn and H. Harari, Nucl. Phys. {\bf B176},  135  (1980).

\bibitem{kApp} For similar treatments, see J.~M. Soares and L.
  Wolfenstein, Phys. Rev. D {\bf 47}, 1021 (1993); G. B{\'{e}}langer
  and C.~Q. Geng, Phys. Rev. D {\bf 43}, 140 (1991).

\bibitem{wolf83} L. Wolfenstein, Phys. Rev. Lett. {\bf 51}, 1945
  (1983).

\bibitem{alta88} G. Altarelli and P.~J. Franzini, Z. Phys. C {\bf 37},
  271 (1988).

\bibitem{kim90} C.~S. Kim, J.~L. Rosner, and C.~P. Yuan, Phys. Rev. D
  {\bf 42}, 96 (1990).

\bibitem{pasc89} E.~A. Paschos and U. T{\"{u}}rke, Phys. Rep. {\bf
    178}, 145 (1989).

\bibitem{bela91} G. B{\'{e}}langer and C.~Q. Geng, Phys. Rev. D {\bf
    43}, 140 (1991).

\bibitem{buch94} G. Buchalla and A.~J. Buras, Nucl. Phys. {\bf B412},
  106 (1994).

\end{references}
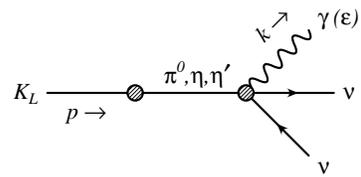
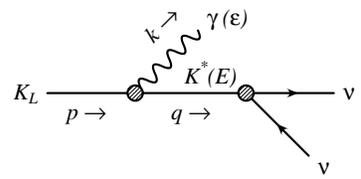
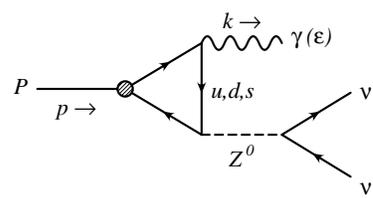
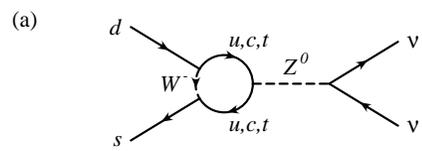
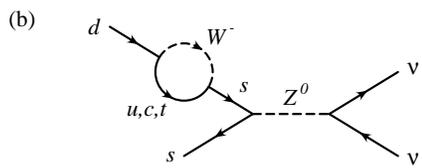
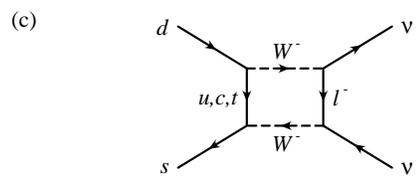
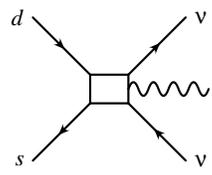
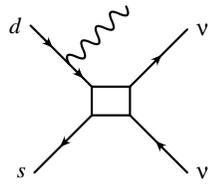
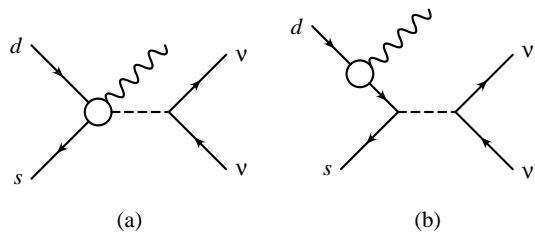
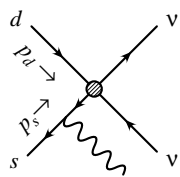
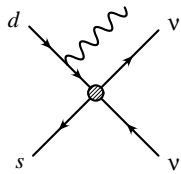
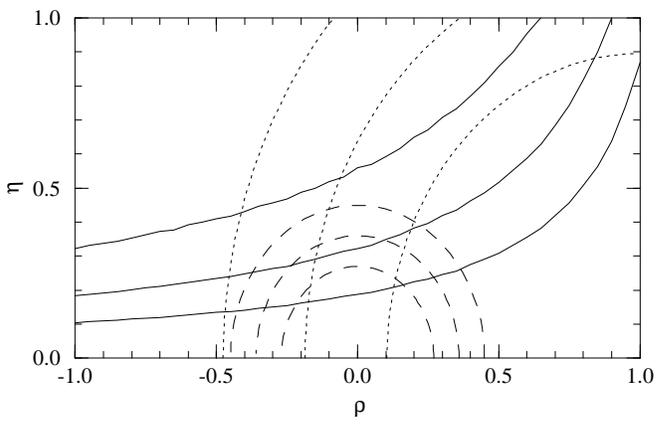
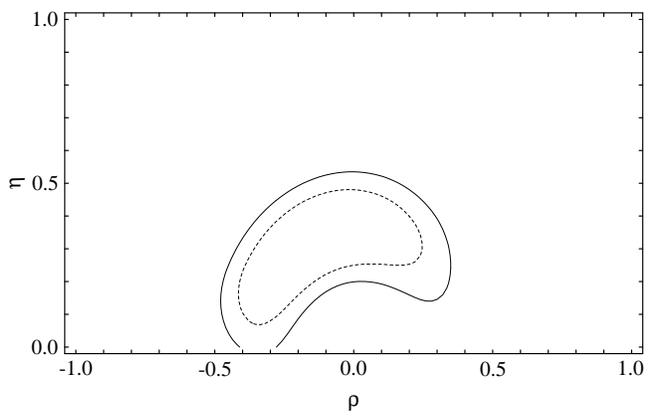
\end{document}